\documentclass[fleqn,usenatbib]{mnras}

\usepackage{newtxtext,newtxmath}

\usepackage[T1]{fontenc}
\usepackage{ae,aecompl}

\usepackage{graphicx}	
\usepackage{amsmath}	

\begin{document}

\title[FRB Down-drifting, bandwidth and beaming]{Down-drifting, bandwidth and beaming of fast radio bursts}

\author[Tuntsov, Pen and Walker]{Artem V. Tuntsov$^{1}$\thanks{E-mail: artem.tuntsov@manlyastrophysics.org}, Ue-Li Pen$^{2}\thanks{E-mail: pen@cita.utoronto.ca}$ and Mark A. Walker$^{1}$\thanks{E-mail: mark.walker@manlyastrophysics.org}\\
\\
$^{1}$Manly Astrophysics, 15/41-42 East Esplanade, Manly, NSW 2095, Australia\\
$^{2}$Canadian Institute for Theoretical Astrophysics, University of Toronto, 60 Saint George Street, Toronto, ON M5S 3H8, Canada}

\date{\today}

\maketitle
\label{firstpage}
\begin{abstract}
'Downwards drifting' structure observed in fast radio bursts (FRBs) could naturally arise from a screen occurring after a small initial dispersion measure region. The screen imprints temporally sharp but broadband structure on the pulse that has already been dispersed, and the `structure-maximising' bulk of the dispersion measure is acquired further along the path to observer. If so, scaling of the drift rate of repeating FRBs with frequency suggests that the emission is beamed -- out of our line of sight -- with the circum-burst plasma deflecting the beam towards us. This in turn explains the observed limited, variable bandwidths of bursts despite broadband nature of the underlying emission. We summarise the geometric constraints on this simple model delivering much of the observed FRB spectro-temporal phenomenology, present generic predictions and discuss possible nature of modulation.
\end{abstract}

\section{Introduction}

The time-frequency plots of fast radio bursts show a common pattern of downwards drifting structure, which has no generally agreed-on explanation to date \citep{cordes2017, wang2019, lyutikov2019, margalit2019, 2020MNRAS.498.4936R}. In particular, burst dynamic spectra of the original repeating FRB 121102, when dedispersed to a common dispersion measure (DM) value that maximises temporal structure, often display several bright patches which have similar spectral width but the central frequency decreasing from the first to subsequent patches; the patches are separated by temporally narrow broadband `shadows' \citep{michilli2018, gajjar2018, zhang2018, josephy2019}. Due to this drift, DM values estimated by maximising the signal-to-noise ratio of the projected pulse are higher than the `structure-maximising' DM. Remarkably, the former values differ from burst to burst whereas the latter is common to all \citep{hessels2019}. Similar downwards drifts are observed in other repeaters \citep{amiri2019, andersen2019, 2021arXiv210604356P} and sharp broadband `shadows' are seen in some non-repeating FRBs where data allows for coherent dedispersion of the signal \citep{farah2018}.

In this short paper we propose modulation near the source as a possible mechanism to generically cause this structure. We interpret the downwards drift simply as an extra dispersion of the burst due to the material between the source and the modulating screen -- including and, possibly, constituting the screen itself. This plasma is presumed to be part of the dynamic environment in the immediate vicinity of the source and thus the extra DM is naturally variable. The structure-maximising DM, on the contrary, is due to electrons of the more mundane regions in the host Galaxy, intergalactic medium and the Milky Way along the line of sight. This is not expected to change on short time scales and thus the DM of the sharp structure is constant.

If the circum-burst plasma is accepted as the origin of the downwards frequency drifts, the observed scaling of the drift rate from band to band allows the deflection angle due to the extra plasma to be the same at all frequencies. A natural interpretation is then that the burst emission is beamed -- away from us -- and the bursts are only seen when bent into our line of sight by an appropriate plasma 'prism'. The deflection angle of every such prism is very chromatic, effectively `scanning' the beam pattern, along one angle, as the frequency changes within the band; away from the optimal frequency the deflected line of sight falls outside of the beam, explaining the limited bandwidth of bursts, which can therefore be intrinsically broadband. Interestingly, for plasma prisms, the sharply defined temporal features in the burst dynamic spectra map to certain points on the screen independent of the frequency; thus, the broadband `shadows' between the burst 'sub-pulses' may be a sign of localised obscuring structures in the circum-burst environment swept over by the beam.

In Section~\ref{section:model} we present, in a few stages of approximation, the model sketched above; it is summarised in Figure~\ref{figure:setup} and further in Figures~\ref{figure:prisms} and~\ref{figure:chromaticity}, which, along with their captions, should give enough of an idea of our work to a busy reader. Section~\ref{section:quantitative} attempts to quantify the physical properties of the circum-burst plasma and typical beaming angle, arguing in favour of relatively wide beams, and interprets a selection of observed FRB trends in properties in terms of our model. We conclude with a brief discussion of modulation mechanisms in Section~\ref{section:discussion}.

\section{Model}
\label{section:model}

\subsection{Sad trombone of Dispersion and modulation}

\begin{figure*}
\centering
\includegraphics[width=1.0\linewidth, angle=0]{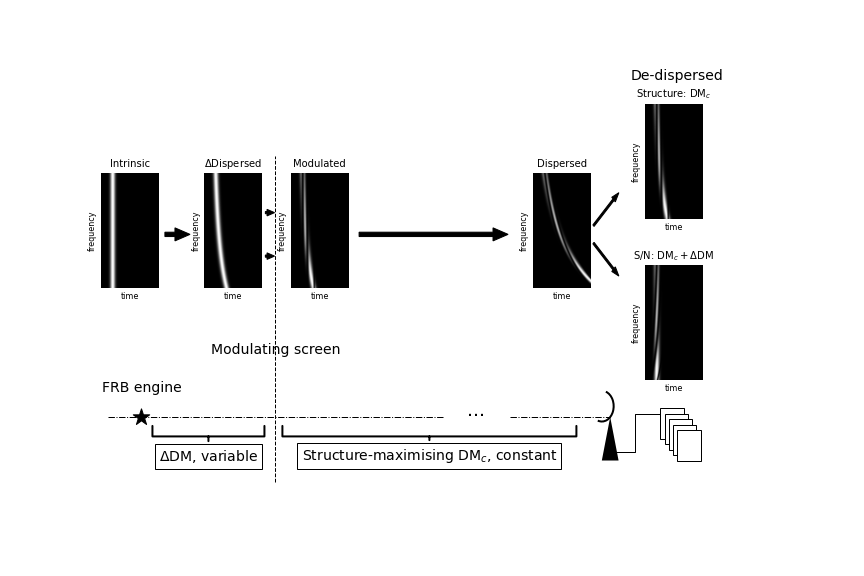}
\caption{
FRB propagation geometry. The first region from the FRB engine labelled $\Delta\mathrm{DM}$ is variable in space and/or time. This is followed by a modulation screen, which can be absorbing or lensing.  The remaining, constant (on short time scales) $\mathrm{DM}$ aggregates further material in the vicinity of the FRB ({\it e.g.}, supernova remnant, photo-dissociation region), its host galaxy, the intergalactic medium and through the Milky Way. The top of figure illustrates changes to the dynamic spectrum as it propagates in this set-up neglecting frequency modulation due to scintillation effects.
}
\label{figure:setup}
\end{figure*}

Our model of the formation of downwards drifting bands is sketched in Figure~\ref{figure:setup}. We propose three stages for this process: a plasma environment near the source that contributes a small $\Delta\mathrm{DM}$, surrounded by a region of intermittent broadband modulation, followed by a long path external to the screen which dominates the $\mathrm{DM}$. The initial small $\Delta\mathrm{DM}$ region is very close to the engine, and can be time variable. This region is also assumed to be inhomogeneous, with varying $\Delta\mathrm{DM}$ along different ray paths. We will stay within the geometric optics limits in this paper.

The mapping from the FRB engine to the screen is, to the first approximation, that of a dispersive delay, with the frequency $\nu$ unchanged (we consider screen on the whole stationary and neglect any relativistic effects) and the intensity density $I$ unaffected apart from the delay:
\begin{eqnarray}
I_e\to I_s(t_s, \nu)=I_e(t_e, \nu), \\
t_e\to t_s=t_e+\frac{cr_e\Delta\mathrm{DM}}{2\pi\nu^2}.\label{dispersion}
\end{eqnarray}
At the screen this dispersively, depending on $\nu$, delayed signal is modulated by a sharp temporal filter $S(t_s)$ -- {\it e.g.}, one or more distinct drop-outs, of widths $w_i$ at positions $t_i$ -- that is broadband, independent of the frequency:
\begin{eqnarray}
I_s\to I_m(t_s, \nu)=I_s(t_s,\nu)S(t_s)=I_e\left[t_e(t_s), \nu\right] S(t_s)\\\nonumber
=I_e\left(t_s-\frac{cr_e\Delta\mathrm{DM}}{2\pi\nu^2}, \nu\right) S(t_s).
\end{eqnarray}
Because the signal arrives at the screen already dispersed, this naturally leads to a 'sad trombone' structure with 'sub-bursts' that arrive later having lower frequencies. The illustration in Figure~\ref{figure:setup} uses a filter with three Gaussian shadows
\begin{eqnarray}\nonumber
S(t_s)=1-\sum\limits_{i=\overline{1,3}}\exp\left[-\frac{(t_s-t_i)^2}{2w_i^2}\right]
\end{eqnarray}
of widths $w_i$ that are well separated, $w_i\ll|t_i-t_{i'}|$, and slightly narrower than the original burst $I_e(t_e)$, also modeled as a broadband Gaussian.

The emission is then further dispersively delayed by a (much larger) contribution to the $\mathrm{DM}_c$ between the screen and the observer
\begin{eqnarray}
t_s\to t_o=t_s+\frac{cr_e\mathrm{DM}_c}{2\pi\nu^2}
\end{eqnarray}
so that
\begin{eqnarray}\label{observed}
I_m\to I_o(t_o, \nu) = I_m(t_s, \nu)=I_s[t_s(t_o), \nu] S[t_s(t_o)]\hspace{.5cm}\\\nonumber
=I_e\left[t_o-\frac{cr_e(\Delta\mathrm{DM}+\mathrm{DM}_c)}{2\pi\nu^2}, \nu\right] S\left(t_o-\frac{cr_e\mathrm{DM}_c)}{2\pi\nu^2}\right)
\end{eqnarray}
where we again neglected Doppler/relativistic ({\it e.g.}, cosmological expansion) effects; likewise, this leg of the journey would often imprint frequency modulation on the signal due to interstellar/intergalactic scintillation which we ignore along with any scattering tail. This further dispersion does not change the qualitative 'sad trombone' structure imprinted by the modulation on a pre-dispersed burst.

The recorded data $I_o(t_o, \nu)$ is customarily dedispersed by realigning the channels depending on frequency, $t_o\to\tilde{t}_o=t_0-cr_e\mathrm{DM}/2\pi\nu^2$ and the last line of~(\ref{observed}) makes this dedispersion ambiguous. The second factor suggests using a dispersion measure of $\mathrm{DM}_c$ which would maximise the structure imprinted by the sharp filter $S(t_s)$. Alternatively, one can use the sum $\mathrm{DM}_c+\Delta\mathrm{DM}$, as suggested by the first factor, to maximise the $\int\mathrm{d}\nu\, I(\tilde{t}_o, \nu)$ signal-to-noise, to which the sharp shadows in the filter do not contribute. The alternative is illustrated in the rightmost part of Figure~\ref{figure:setup}. Similar ambiguity arises if there is more than one modulating screen along the line of sight, with further $\Delta\mathrm{DM}$ acquired from one to another -- leading to more complex spectro-temporal structure with possibly intersecting patterns.

\subsection{Misaligned beam from drift rates}

In the presented model, the 'sad trombone' effect is just a time delay that depends on the frequency. Traditionally though, it has  been reported as the rate of `frequency drift with time' in waterfall plots dedispersed to the structure-optimising value $\mathrm{DM}_c$ of the dispersion measure. Locally, within a narrow band, these are just inverse of each other, so the drift rate due to a pre-disperson by $\Delta\mathrm{DM}$ prior to the modulation is
\begin{eqnarray}\label{driftrate}
\frac{\partial\nu}{\partial t_s}=\left(\frac{\partial t_s}{\partial\nu}\right)^{-1}=-\frac{\pi\nu^3}{cr_e\Delta\mathrm{DM}}.
\end{eqnarray}
The prediction of our model would then be that, for a burst measured over a wide bandwidth, the drift rate would be proportional to the frequency cubed.

We are not aware of such measurement for a single wide-band burst (or an array of sub-bursts) but the drift rates measured for \emph{different} bursts from the repeating FRB~121102 in different bands appear to correlate with the band centre frequency in a linear proportion \citep{josephy2019}:
\begin{eqnarray}\label{ratescaling}
\left\langle\frac{\partial\nu}{\partial t_s}\right\rangle_i\propto\nu_i.
\end{eqnarray}
To reconcile this scaling with the prediction~(\ref{driftrate}) of the model, one would need the extra column density between the screen and the FRB engine to correlate -- from one burst to another -- with the frequency squared:
\begin{eqnarray}\label{dmscaling}
\left\langle\Delta\mathrm{DM}\right\rangle\propto\nu^2,
\end{eqnarray}
where the angled brackets reiterate that this relationship is meant to be not a deterministic prescription but rather a statistical correlation followed by an ensemble of bursts from the same source, possibly with some variation -- as is clearly present in the relation~(\ref{ratescaling}) observed by \cite{josephy2019}.

\begin{figure}
\centering
\includegraphics[width=1.0\linewidth, angle=0]{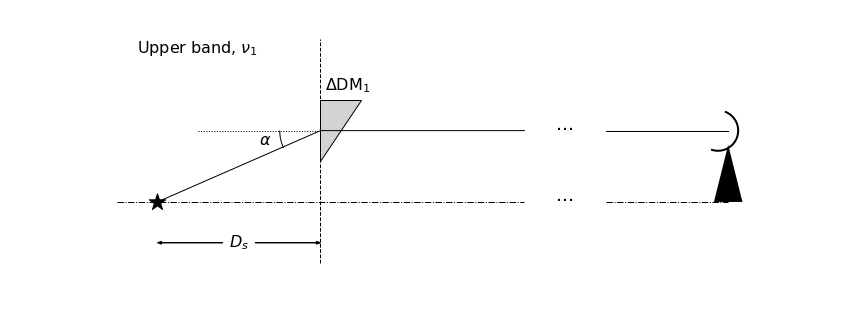}
\includegraphics[width=1.0\linewidth, angle=0]{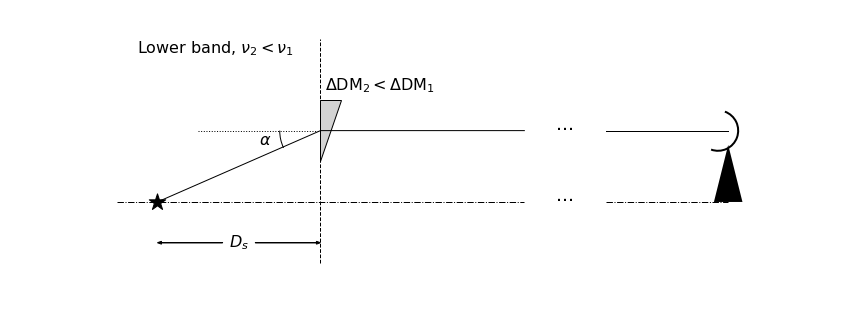}
\caption{Deflection of beamed radiation into our line of sight by a plasma prism. To keep the deflection angle constant, as set by the misalignment of the beam from our line of sight, the density gradient should scale in proportion to frequency squared, which agrees with the observed scaling of down-drift rates with frequency.
}
\label{figure:prisms}
\end{figure}

And indeed there is a setting where such correlation is expected. If the intrinsic FRB emission is beamed \citep{2017MNRAS.467L..96K}, and so that our line of sight is outside the beam, for the burst to be detected its emission needs to be deflected towards us by the misalignment angle, $\alpha$. For a plasma prism -- {\it i.e.}, a gradient of the plasma column density across the line of sight, the deflection angle is\footnote{The minus sign reflects that plasma advances the wavefront rather than retarding it; a plasma prism bends rays towards its apex, not base, as pictured in Figure~\ref{figure:prisms}.},
\begin{eqnarray}\label{deflection}
\alpha=-\frac{c^2r_e}{2\pi\nu^2}\nabla\Delta\mathrm{DM},
\end{eqnarray}
-- {\it i.e.}, proportional to $\nu^{-2}$, so to keep the deflection fixed at the value of the misalignment between the beam and our line of sight one would need the gradient to scale as
\begin{eqnarray}\label{gradientscaling}
\nabla\Delta\mathrm{DM}\propto\nu^2;
\end{eqnarray}
note that there are not angled brackets in this equation. For~(\ref{gradientscaling}) to be equivalent to~(\ref{dmscaling}), the baseline $\Delta x$ of the derivative $\nabla\Delta\mathrm{DM}=\Delta\mathrm{DM}/\Delta x$ needs to be, on average, independent of frequency -- {\it e.g.}, set by the geometric constraints on the screen. When we do see a burst, it has been deflected by the required angle, which needs a stronger plasma density gradient at higher frequency. 

Thus, our updated model, as shown in Figure~\ref{figure:prisms}, upgrades a plasma slab in front of the modulating screen to a plasma prism -- or, more precisely, different prisms at different bands all having the same deflection angle. We presume that rather than being fixed objects such prisms form and disappear at various positions in the turbulent circum-engine plasma medium, so that the bursts are seen when their deflection angle matches the required misalignment value. Note, that in the updated model, the rays follow broken rather than straight lines to the source, which adds to the dispersive term in~(\ref{dispersion}) a geometric contribution to the time delay:
\begin{eqnarray}
\Delta t_\mathrm{g}=\frac{D_s}{c}\left(\frac1{\cos\alpha}-1\right),
\end{eqnarray}
where $D_s$ is the distance from the engine to the prism along the line of sight. For this contribution to not affect the frequency scaling relied on in the previous and next subsection, it needs to be (statistically) independent of frequency (which would apply to $D_s$ as $\alpha$ is frequency-independent in the model), or small compared to the dispersive term. 

A simple but important point is that modulation itself is \emph{not} required for deflection; in our model, all observed bursts are deflected by the same angle, including those not split into subpulses.

\subsection{Limited bandwidth by prism chromaticity}

\begin{figure}
\centering
\includegraphics[width=1.0\linewidth, angle=0]{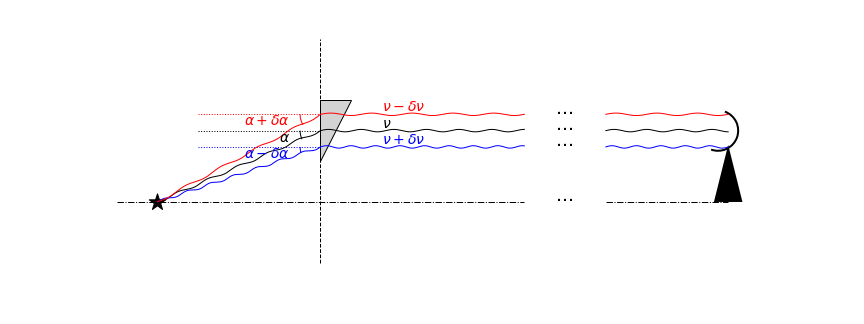}
\includegraphics[width=1.0\linewidth, angle=0]{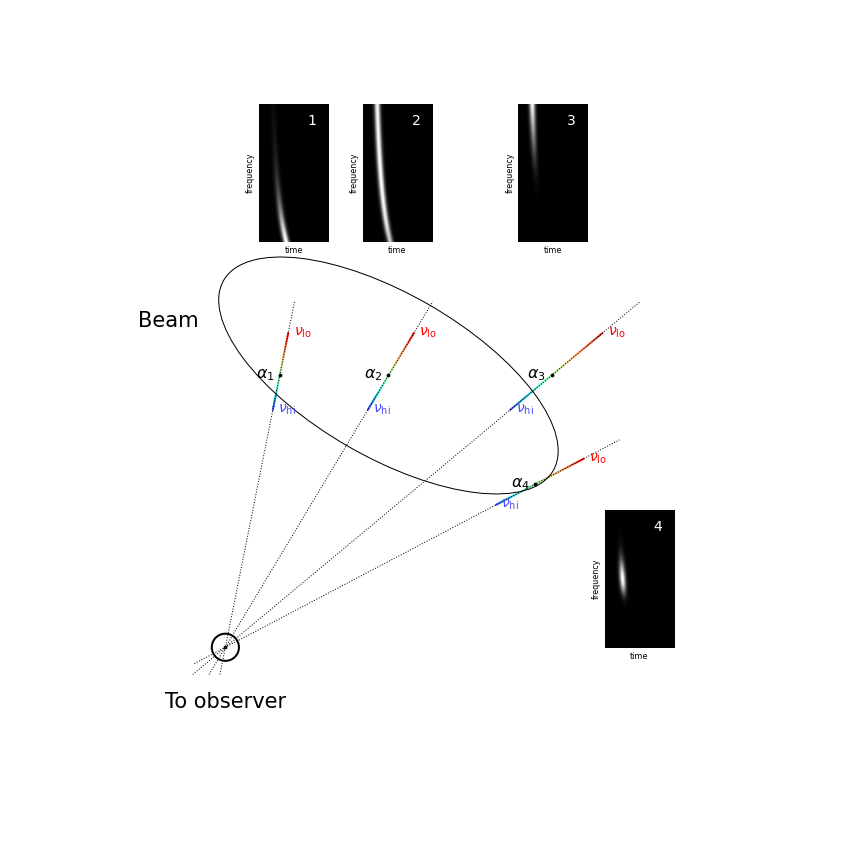}
\includegraphics[width=0.45\linewidth, angle=0]{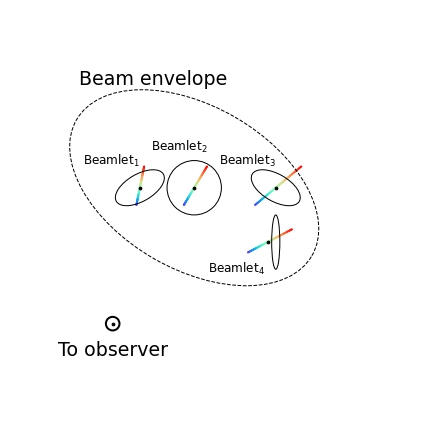}
\includegraphics[width=0.45\linewidth, angle=0]{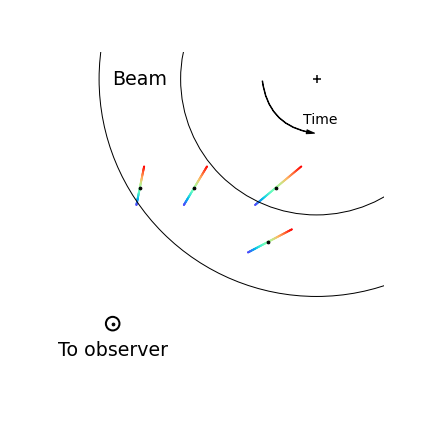}
\caption{Chromatic scanning of the beam. {\it Top panel.} For a given plasma prism, the deflection angle depends strongly on the frequency, and thereby changing frequency effectively 'scans' the emission beam in the dispersion direction. To avoid confusion when interpreting this diagram in the lensing context, please note that in the prism approximation, at a given frequency, the deflection angle is independent of the position in the prism, so there is no spatial convergence of rays in the prism approximation. Rather, it is the position of the deflection point that changes with frequency, as shown in the plot. {\it Middle panel.}  Formation of band-limited bursts depending on the position of the dispersive 'cut' in the beam. Dynamic spectrum illustration cut-outs assume a Gaussian beam. {\it Bottom panels} illustrate complications in reconstructing the beam profile from dispersive cuts if the beam is broad, is only defined as a statistical envelope of individual `beamlets' varying from burst to burst, has a complicated shape and/or time variable. }
\label{figure:chromaticity}
\end{figure}

Although the prisms required to send the beamed burst towards us differ from band to band (and, possible, from one occasion to the next within a band), the converse is true for a given realisation of the prism: as the frequency is changed, so does the deflection angle, and so tuning our receiver through the burst bandwidth effectively 'scans' the beam profile, as shown in Figure~\ref{figure:chromaticity}. This opens the possibility that the band-limited character of the observed bursts can actually reflect its angular beaming and the chromaticity of the prism that facilitates detection of the burst; the underlying emission mechanism might well be broadband.

Specifically, in the case of beamed emission it will be convenient to write the lens equation as
\begin{eqnarray}\label{lensdef}
\pmb{\theta}-\pmb{\alpha}=\pmb{\eta},
\end{eqnarray}
where $\pmb{\theta}$ labels the beam pattern, $\pmb{\alpha}$ is the deflection angle introduced above and $\pmb{\eta}$ is the direction to the fiducial observer. We will choose the latter as the origin of the coordinate system, $\eta\equiv0$, which renders  $\pmb{\theta}=\pmb{\alpha}$ but nevertheless keep separate variables for the coordinate in the beam, $\pmb{\theta}$ and the prism deflection angle -- for clarity and in reserve for upgrading the prism to a more general lens in the end of this section. To the extent to which the plasma can be represented as a prism though, the deflection angle only depends on the frequency, so that
\begin{eqnarray}\label{prismequation}
\pmb{\theta}=\pmb{\alpha}_0\frac{\nu_0^2}{\nu^2}
\end{eqnarray}
where $\alpha_0$ is the deflection angle at a reference frequency $\nu_0$.

Then, the formation of band-limited bursts from broadband emission with an angular beam pattern of $B(\pmb{\theta})$\footnote{$B(\pmb{\theta})$ could in principle depend on the frequency explicitly but that would be an unnecessary complication at this stage.} can be described as:
\begin{eqnarray}\label{beamscan}
\nu\to\pmb{\theta}={\pmb\alpha}_0\frac{\nu_0^2}{\nu^2}, \hspace{.3cm} B\to I\left(\nu\right)=B\left[\pmb{\theta}(\nu)\right]=B\left({\pmb\alpha}_0\frac{\nu_0^2}{\nu^2}\right).
\end{eqnarray}
As frequency changes, a `cut' $\pmb{\theta}(\nu)$ can enter the beam, trace it along the cut and then, possibly, leave while $\nu$ is still within the observed bandwidth. Alternatively, the cut may fit entirely within the band (or never make it inside -- though in this case a burst would not be observed); a variety of cuts at different orientations and positions can be probed in different bursts as illustrated in the middle panel of Figure~\ref{figure:chromaticity}.

Conversely, cuts through a broadband beam pattern $B(\pmb{\theta})$, adjacent and parallel to reference position $\pmb{\alpha}_0$ associated with a burst, can be read off the detected bursts dynamic spectra via the inverse mapping:
\begin{eqnarray}
I\to B(\pmb{\theta})=I\left[\frac{\nu_0\,\alpha_0}{\sqrt{\left({\pmb\alpha_0}\,\pmb{\theta}\right)}}\right], \hspace{.2cm} B(\theta_\parallel, \theta_\perp)=I\left(\nu_0\sqrt{\frac{\alpha_0}{\theta_\parallel}}\right).
\end{eqnarray}
 It is not clear how one can determine the reference positions $\pmb{\alpha}_i$ of these cuts for a broad two-dimensional pattern that permit burst detection for a wide range of $\pmb{\alpha}_0$. Further complications, as illustrated in the bottom panels of Figure~\ref{figure:chromaticity}, are possible. The beam that facilitates observation of bursts may only be defined statistically as an envelope of individual 'beamlets' varying from burst to burst -- or/and the beam is (beamlets are) time-dependent from burst to burst (within each burst), as in the `lighthouse sweep' pattern invoked for pulsars.

The next level of refinement to the model would allow the plasma column gradient to vary across the screen, $\nabla\Delta\mathrm{DM}({\bf x})$, replacing~(\ref{lensdef}) with
\begin{eqnarray}\label{lenseq}
\pmb{\theta}-\pmb{\alpha}_0\left(D_s\pmb{\theta}\right)\frac{\nu_0^2}{\nu^2}=\pmb{\eta},
\end{eqnarray}
where the rays are mapped from the engine to the screen via ${\bf x}=D_s\pmb{\theta}$ and an explicit function
\begin{eqnarray}
\pmb{\alpha}_0({\bf x})=-\frac{c^2 r_e}{2\pi\nu_0^2}\nabla\Delta\mathrm{DM}({\bf x})
\end{eqnarray}
needs to be specified. We are still free to choose $\pmb{\eta}\equiv0$ as in~(\ref{prismequation}), however as the deflection is now $\pmb{\theta}$-dependent, the observed dynamic spectrum needs to be corrected for the magnification factor\footnote{Note that the position(s) $\pmb{\theta}$ at which the Jacobian is to be computed is itself frequency-dependent, thus the explicit $\nu$ dependence in~(\ref{magnification}) is not sufficient to establish the frequency dependence of the function.}:
\begin{eqnarray}\label{magnification}
\mu=\left|\det \frac{\mathcal{D}(\pmb{\theta})}{\mathcal{D}(\pmb{\eta})}\right|=\left|\det\left(\mathrm{1}-\frac{\nu_0^2}{\nu^2}\partial_{\pmb{\theta}}\pmb{\alpha}_0\right)\right|^{-1}
\end{eqnarray}
due to convergence of the rays at a given frequency, replacing the beam-watefall mapping part of~(\ref{beamscan}) with:
\begin{eqnarray}\label{beamscanlens}
B\to I\left(\nu\right)=\mu[\pmb{\theta}(\nu)]B\left[\pmb{\theta}(\nu)\right].
\end{eqnarray}
The frequency-angle remapping part formally remains the same but the $\pmb{\theta}(\nu)$ there now refers to the solution of
\begin{eqnarray}\label{prismequationlens}
\pmb{\theta}=\pmb{\alpha}_0(D_s\pmb{\theta})\frac{\nu_0^2}{\nu^2}
\end{eqnarray}
parameterised by $\nu$ rather than an explicit function of $\nu$ parameterised by $\pmb{\alpha}_0$.

The analysis of the full lens equation, with the associated magnification, solution multiplicity and caustic structure, all morphing with frequency, is a profoundly more complex problem \citep{cordes2017}, even with simple models for $\Delta\mathrm{DM}({\bf x})$. We want to emphasise, however, that a far more restricted version of this problem studied in this paper is capable of explaining some of the salient spectro-temporal phenomenology of FRBs, including frequency down-drifting, band-limited character of emission without resorting to exotic emission physics. It is worth stressing that the substantial distinction of this scenario from the plasma lensing model of \cite{cordes2017} is that the drifting itself is due to dispersion and not lensing; the latter might still be responsible for modulation of the burst at the screen. 

\section{Physical scales}
\label{section:quantitative}
The discussion so far has mostly been qualitative as much of the phenomena are geometric in origin. In this section we get more specific by interpreting the observed properties of fast radio bursts within our model and inferring physical properties this interpretation implies, which we check, where possible, against independent constraints. We will base the comparison around the properties of the original repeater, FRB~121102 \citep{2016Natur.531..202S, 2014ApJ...790..101S}, which, arguably, remains the most extensively studied burst source so far, and around the frequency of $1\,\mathrm{GHz}$ where majority of the detections are made.

The plasma volume density corresponding to a plasma frequency $\nu_p$ of $1\,\mathrm{GHz}$ is $n_e^\mathrm{max}\approx10^{10}\,\mathrm{cm}^{-3}$, which is the highest value along the ray between the FRB engine and the screen can go through without being absorbed. The drift rates reported at this frequency correspond to $\Delta\mathrm{DM}\sim1\,\mathrm{pc}\,\mathrm{cm}^{-3}$ (which can also be inferred from the reported spread of the S/N-maximising values, to which downwards drifting contributes -- though that is not an independent measurement within our model). Hence, for the plasma frequency to be comfortably below $1\,\mathrm{GHz}$, the extra DM needs to be accumulated over a distance of $D_s\gtrsim 10^9\,\mathrm{cm}$; it could be much higher, of course. Observations at lower frequency constrain $n_e^\mathrm{max}$ further but the $\Delta DM$ required to fit the drift rates at this frequency is also much reduced (and in the same proportion), so the constraints on $D_s$ is not changed. At such distance, the Fresnel radius is greater than $10^5\,\mathrm{cm}$ at $1\,\mathrm{GHz}$ setting the minimum linear size on plasma structures that could function as prisms or opaque objects able to create broadband 'shadows'. 

The angular scales of the model, most notably the deflection angle, are not directly constrained by the data. The deflection angle due to a gradient equal to the $n_e^\mathrm{max}$ is $\nu_p^2/2\nu^2$ so can be $\mathcal{O}(1)$ close to the plasma frequency. Moreover, a gradient of the column through a smoothly bound overdensity can take an arbitrarily high value sufficiently close to the projected edge, and our model suggests that the observed bursts are selected by them being deflected through the required angle, so the argument that the angle be 'typical' does not apply. However, the model allows us to compare the deflection angle to the beam width. As the dispersive cut 'width' is about twice the observed bandwidth, presence of broadband bursts in the sample argues that the beam width is not much narrower than the deflection angle. The latter is set by the misalignment of our line of sight from the beam (envelope) boundary and the likelihood of the misalignment increases with the misalignment angle -- {\it e.g.}, quadratically or linearly for a narrow pencil or fan beam geometry, respectively. The fact that we are not seeing ten to a hundred more FRB sources producing tenth-of-the-observed-bandwidth bursts than there are sources of broadband bursts suggests that either the deflection angles are somehow limited to near the beam width, which is unlikely given $\alpha\propto\nu^{-2}$ for a given column gradient, or that the beams are not narrow. In the latter case, there should be a fraction of repeating FRBs not relying on deflection to get into our line of sight, which will not follow the linear scaling of drift rate with frequency; the rest of the phenomenology might still be present.

\section{Discussion}
\label{section:discussion}

 We propose that the downwards drifts in the dynamic spectra of FRBs dedispersed to `structure-maximising' value of the dispersion measure result because the broadband temporal structure is imprinted \emph{after} the pulse has already travelled through a substantial DM in the immediate vicinity of the source. In the former case, the substantial distinction from the plasma lensing model of \cite{cordes2017} is that the drifting itself is due to dispersion and not lensing; this breaks the time reversal symmetry and explains why only downwards drifting pulses are observed. 

The tentatively observed scaling of the drift rate from band to band then suggests that the initial DM required for the mechanism to operate is proportional to the square of frequency, which is compatible with the demands on plasma lensing of the source beam into our line of sight. The nature of modulation at the screen remains unclear. For a sharp feature on top of the pre-dispersed burst, the screen should block emission quickly, on timescales smaller than the burst duration and extra dispersion from the engine to the screen. The blocking patch should be larger than the projection of the dispersive 'cuts' shown in Figure~\ref{figure:chromaticity} onto the screen. It also needs to be larger than the Fresnel scale -- this requirement applies to all other structures studied in this paper which limits itself to geometric optics. Interestingly, the time delay~(\ref{dispersion}) and the deflection position~(\ref{prismequation}) have the same dependence on frequency -- and thus, if the radiation is blocked through a particular geometrical alignment of the sweeping emission beam with the screen, such alignment at one particular frequency could, in principle, extend over the entire bandwidth. The amplitudes of the two $\propto\nu^{-2}$ effects are essentially independent though, so unless the alignment itself is adaptive ({\it e.g.}, the fractions of the two amplitudes relevant to the alignment change through the duration of the burst), such arrangement appears contrived. 

The physical mechanism of opacity itself is not known. It might simply reflect a particularly dense region along the ray from the engine to the patch where the plasma frequency exceeds the upper edge of the observed band. In pulsar-brown dwarf systems, the complete absorption of radio waves is observed, the nature of which is not understood \citep{1994ApJ...422..304T,2019MNRAS.484.5723L}. Similarly, magnetospheric eclipses have been observed in the double pulsar system \citep{2005ApJ...634.1223L}, whereby the slower B pulsar modulates its companion flux at a variable rate, which is related to rotational and orbital time scales; the millisecond variability in FRBs would then imply a millisecond pulsar companion with an irregular magnetosphere.  Beyond opacity, plasma lensing, observed in a broad range of environments, from the ionosphere to the solar wind, interstellar medium, or near pulsars \citep[e.g.,][]{2018Natur.557..522M} is demonstrably capable of producing sharp temporal features, and gravitational \citep[e.g.,][]{2020ApJ...900..122S} or gas, refractive \citep{draine1998} lensing are less frequency-dependent that their plasma counterpart. Possible role for non-linear wave propagation in the formation of spatio-temporal structure has been suggested (\citealt{2019arXiv191208150G, 2020ApJ...892L..10Y, 2020MNRAS.494..876D, 2021MNRAS.500..272S} -- though \citealt{2020arXiv200109210L}).

Recently, burst pairs with central frequencies increasing with time have been detected \citep{2020Natur.587...54C}, albeit in a `Galactic' FRB \citep{2020Natur.587...59B}. The dynamic spectra of such pairs do look distinctly different from the more familiar sequences of downwards-drifting pulses in more ways than just the direction of the drift. We therefore believe that they represent a different, possibly physical phenomenon, unrelated to the \emph{apparent} downwards drifting due to the modulation of a pre-dispersed burst studied in this paper.

There is a number of consistency relations that the model can be tested against. The curvature of the electron column density cannot exceed a value of $\sim\lambda^2 D_s r_e$ to avoid significant demagnification of bursts. For repeaters, one could also test the relationship between $\Delta DM$ observed for different bursts and at different bands and the variation in the rotation measure (D.Z. Li, private communication). One particularly intriguing test involves the slope of the drift rate {\it vs} band frequency relation, which, in the present model, is just the inverse of the geometric time delay between the direct and deflected rays; this value is surprisingly similar from one repeater to the next.

\section*{Acknowledgements}
AVT thanks the organisers and participants of `FRB2018' (Swinburne Uni), 'Gravity meets Plasma' (SWIfAR) and 'Scintillometry 2019' (MPIfRA) workshops and is grateful to Shivani Bhandari, Adam Deller, Xinzhong Er, Maxim Pshirkov, Laura Spitler, Wei-Yang Wang and Bing Zhang for short stimulating discussions and to Keith Bannister for more extensive but no less intense ones.


\bibliographystyle{mnras}
\bibliography{marching}

\label{lastpage}
\end{document}